\documentclass[journal]{IEEEtran}

\usepackage{textcomp}
\usepackage{amsmath,amssymb,amsfonts}
\usepackage{algorithmic}
\usepackage{graphicx}
\graphicspath{{./Figures/}}
\DeclareGraphicsExtensions{.pdf,.jpeg,.png}
\usepackage{array}
\usepackage{wrapfig}
\usepackage{subcaption}
\usepackage{stfloats}
\usepackage{verbatim}
\usepackage{rotating}
\usepackage{comment}
\usepackage[style=ieee, backend=bibtex]{biblatex}
\begin{document}
%
\title{An Active Electromagnetic 3D Surface Cloak}
%
%
%

\author{Paris Ang and George V. Eleftheriades
\thanks{This work has been supported by the Natural Sciences and Engineering Research Council of Canada (NSERC), the Ontario Centers of Excellence (OCE), the Department of National Defence Canada (DND), and AUG Signals.}
\thanks{The authors are with the Edward S. Rogers Sr. Department of Electrical and Computer Engineering, University of Toronto, Toronto, ON M5S 3G4, Canada	(e-mail: paris.ang@mail.utoronto.ca; gelefth@waves.utoronto.ca).}}

\maketitle

\begin{abstract}
An active interior cloak is composed of a conformal array of radiating sources surrounding a target object. With proper configuration, this array radiates a field discontinuity which cancels out any scattering which occurs when the object is illuminated by an impinging electromagnetic wave. Unlike many other cloaking methods, an interior cloak can be completely constructed using conventional components while its active nature allows it to circumvent passivity-based constraints. This enables an interior cloak to hide an object of virtually any size, shape, and material composition as well as accommodate multiple frequencies and polarizations. Despite its potential, this approach has only been demonstrated on 2D and quasi-2D targets to date. As such, this paper presents the first 3D implementation of an interior cloak. Here, the design of a cloak array capable of hiding a conductive, circular cylinder in free-space is first presented. Cloaking performance is then validated via full-wave simulations and an experimental demonstration in which the device successfully suppresses scattering when the target is illuminated by a ridged horn antenna at 1.2 GHz.
\end{abstract}


%
\IEEEpeerreviewmaketitle

\section{Introduction}
Electromagnetic cloaking refers to techniques and technologies which prevent an object from scattering an impinging electromagnetic wave, usually for the purpose of evading detection. Although the concept of such ``invisibility cloaks" has long existed within the realm of fiction, the advent of metamaterials and metasurfaces have made the realization of real-world devices possible. Typically, the majority of cloaking devices reported so far operate by either redirecting the incident wavefront around an object \cite{Schurig2006, Kim2016,Ma2013,Chen2012}, or by cancelling out scattered fields arising from target/wavefront interaction \cite{Rainwater2012,Vitiello2016,Qin2018}. Furthermore most cloaks are passive in nature, allowing for operationally simple devices but at the cost of fundamental design and performance limitations. Specifically, the design and implementation of passive devices generally require compromises on operational bandwidth, directionality, polarization, as well as the cloak and target's permissible size, geometry, and material composition \cite{Fleury2015,Monticone2013}.

One method of relaxing or mitigating these constraints is to incorporate active components into cloak design. A popular approach is to augment existing passive cloaks with active components and networks \cite{Zeng2014, Hosseininejad2019, Meng2017, Yang2016b}. This results in hybrid devices with expanded capabilities \cite{Chen2013}, such as reconfigurability \cite{Liu2014, Farhat2013} and unidirectional reflectivity \cite{Zhu2013, Zhu2014}, but function similarly to their passive counterparts. Alternatively, there are solely active methods of cloaking where an array of radiating sources are configured to produce a field discontinuity which prevents scattering \cite{Miller2006, Vasquez2009, Selvanayagam2012}. Due to their active nature, such cloaks are theoretically capable of hiding an object of any size, shape, and material composition while accommodating an incident wave of any geometry, polarization, or frequency range. Additionally, these radiating components can take the form of regular antennas, enabling a cloak to be constructed completely using commercial off-the-shelf components.

These purely active cloaks can assume either an ``exterior" or ``interior" configuration. Exterior cloaks employ their sources to cancel out the incident fields within an area, creating a zero-field region \cite{Vasquez2009, Zheng2010b, Du2012, Wong2018}. Consequently, any object(s) placed within this space is then isolated from the impinging wavefront; analogous to the aforementioned re-directive cloaks. Conversely, an ``interior" cloak can be considered cancellation based. Here, a target is surrounded by a conformal array \cite{Miller2006, Selvanayagam2012} of sources which are configured to cancel out the scattered fields following target/wavefront interaction. Interior cloaking was first experimentally demonstrated in 2013 \cite{Selvanayagam2013} in which an array of loop antennas was used to cloak a quasi-2D, metallic, circular cylinder. Since then, recent research has focused on expanding the capabilities and practicality of active cloaking with efforts such as: adapting the cloak for complex, irregular 2D target geometries \cite{Ang2018, Ang2020}, developing low-profile elements to improve array versatility and robustness \cite{Ang2018a, Ang2020}, and developing algorithms and procedures for self-adapting cloak designs \cite{Ang2020a, Ang2021a, Ang2021}.

\begin{figure*}[t]
	\centering
	\includegraphics[width=6.5in]{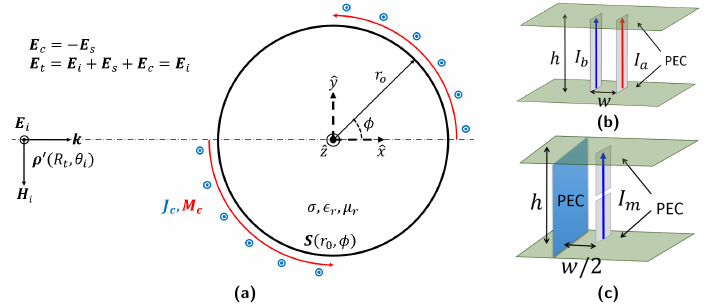}
	\begin{subfigure}{0em}
		\phantomsubcaption{}
		\label{fig:2DLayout}
	\end{subfigure}
	\begin{subfigure}{0em}
		\phantomsubcaption{}
		\label{fig:DefaultHS}
	\end{subfigure}
	\begin{subfigure}{0em}
		\phantomsubcaption{}
		\label{fig:ReducedHS}
	\end{subfigure}
	\caption{(a) Conceptual 2D circular cloaking layout (b) hybrid and (c) simplified equivalence sources \cite{Wong2016}}
	\label{fig:2DLayoutF} 
\end{figure*}

Accommodating real-world target geometries is an essential capability for any practical cloak. Although previous implementations have experimentally verified compatibility with non-uniform scatterers and complex geometric features, like corners and edges, interior cloaking has solely been demonstrated on 2D/quasi-2D targets \cite{Selvanayagam2013, Ang2020}. To address this shortcoming, this publication documents the development and demonstration of a proof-of-concept, active cloak capable of hiding a 3D target under free-space conditions. This is accomplished by extrapolating a previous 2D circular cloak layout into an expanded array capable of hiding a 3D conductive, cylindrical target when illuminated radially at 1.2 GHz by a ridged horn antenna. General active cloaking theory is first presented followed by device development and full-wave simulations for initial validation. After, experimental measurements are then conducted to verify real-world performance.

\section{Active Electromagnetic Cloaking}
The principles behind active cloaking remain universal regardless of incident field and target properties. As such, it is more convenient to describe operation using the 2D cloak problem from Fig. \ref{fig:2DLayoutF}. The layout in Fig. \ref{fig:2DLayout} details a circular target illuminated by an arbitrary incident electromagnetic wave ($\mathbf{E}_{i}$, $\mathbf{H}_{i}$) at some angle of incidence $\theta_{i}$ and, if not planar, originating from radial distance $R_{t}$. The resultant wavefront/target interaction then gives rise to secondary scattered fields inside ($\mathbf{E}_{int}$, $\mathbf{H}_{int}$) and outside ($\mathbf{E}_{s}$, $\mathbf{H}_{s}$) of the target. These fields can be related along the target's surface ($\mathbf{S}\left(r_{0},\phi\right)$) by the following boundary conditions:
\begin{gather}
	\mathbf{\hat{n}}\times\mathbf{E}_{i} + \mathbf{\hat{n}}\times\mathbf{E}_{s} = \mathbf{\hat{n}}\times\mathbf{E}_{int}
	\label{eqn:EBC}\\
	\mathbf{\hat{n}}\times\mathbf{H}_{i} + \mathbf{\hat{n}}\times\mathbf{H}_{s} = \mathbf{\hat{n}}\times\mathbf{H}_{int}
	\label{eqn:HBC}
\end{gather}
where the outward facing normal is represented by $\mathbf{\hat{n}}$. Using the Huygens' equivalence principle \cite{Harrington2001}, allows the scattered fields to be represented as the products of equivalent radiating sources impressed into the target's surface \cite{Selvanayagam2012, Ang2019}; represented by magnetic ($\mathbf{M}_{s}$) and electric ($\mathbf{J}_{s}$) surface current densities:
\begin{gather}
	\mathbf{M}_{s} = -\mathbf{\hat{n}} \times\left(\mathbf{E}_{s}-\mathbf{E}_{int}\right)
	\label{eqn:MS}\\
	\mathbf{J}_{s} = \mathbf{\hat{n}} \times\left(\mathbf{H}_{s}-\mathbf{H}_{int}\right)
	\label{eqn:JS}	
\end{gather}
If physically realized, these sources would radiate a copy of the scattered field distribution, in the absence of the original incident wave. Using this mechanism, it is then possible to create an active cloak by implementing these sources with an added $180^{\circ}$ phase offset ($\mathbf{M}_{c}=-\mathbf{M}_{s}$, $\mathbf{J}_{c}=-\mathbf{J}_{s}$). This creates a phase-inverted version of the scattered field ($\mathbf{E}_{c}, \mathbf{H}_{c}$) which cancels out the natural scattered counterpart when the cloak/target assembly is re-illuminated by the original incident wave. Furthermore, by rearranging and substituting in (\ref{eqn:EBC}) and (\ref{eqn:HBC}), it is possible to reduce (\ref{eqn:MS}) and (\ref{eqn:JS}) to:
\begin{gather}
	\mathbf{M}_{s} = \mathbf{\hat{n}} \times\mathbf{E}_{i}
	\label{eqn:MCurrent}\\
	\mathbf{J}_{s} = -\mathbf{\hat{n}} \times\mathbf{H}_{i}
	\label{eqn:ECurrent}
\end{gather}
This shows that only knowledge of the incident field is required to deduce the scattered field, equivalent sources, and design a cloak \cite{Selvanayagam2012}. Lastly, a conductive target ($\sigma>>0$) shorts out any electric surface source currents ($\mathbf{J}_{c}=0$) and eliminates the formation of any internal electric and time-varying magnetic fields ($\mathbf{E}_{int}=\mathbf{H}_{int}=0$). As such, a cloak designed for this case only requires magnetic sources.

\subsection{Cloaking Elements}
Once deduced, an active cloaking solution can be physically implemented by surrounding the target with a conformal array of $N$ radiating sources. This is accomplished by first discretizing the continuous magnetic ($\mathbf{M}_{c}$) and electric ($\mathbf{J}_{c}$) surface source currents into $N$ dipole moments:

\begin{gather}	
	\boldsymbol{\rho}_{m}=\mathbf{M}_{c}(r_{0},\phi_{n})hl
	\label{eqn:Mdm}\\
	\boldsymbol{\rho}_{e}=\mathbf{J}_{c}(r_{0},\phi_{n})lh
	\label{eqn:Edm}
\end{gather}

where $h$ designates the target height, $l$ represents the spacing between element centers, while $\mathbf{M}_{c}(r_{0},\phi_{n})$ and $\mathbf{J}_{c}(r_{0},\phi_{n})$ represent the magnetic and electric surface currents at the location of a specific element ($n$). Each of these dipole moments can then be produced by appropriately configured (``weighted") magnetic and electric radiating sources. These can take the form of conventional antennas and previous 2D/quasi-2D designs have successfully approximated magnetic moments using loop antennas (of area $A$), fed with a current of: $I_{m}=\boldsymbol{\rho}_{m}/\left(j\omega\mu A\right)$ \cite{Balanis2012b}, as well as patch antennas \cite{Ang2018a}. Likewise, an electric dipole moment can be produced by dipole or monopole antennas weighted with: $I_{e}=\boldsymbol{\rho}_{e}/h$. 

Alternatively, it is possible to simultaneously generate both electric and magnetic moments using hybrid elements such as the ``equivalence source" in Fig. \ref{fig:DefaultHS} \cite{Wong2016}. In this case, a pair of dipole or monopole antennas are weighted with input currents:
\begin{gather}
	I_{a} = 0.5 I_{e} + I_{m}\\
	I_{b} = 0.5 I_{e} - I_{m}
\end{gather}
Here, the even mode created from the two $0.5 I_{e}$ currents acts as a single $I_{e}$ fed electric dipole while a $I_{m}$ weighted current loop (magnetic dipole) is approximated by the odd mode formed by the opposing $\pm I_{m}$ currents.

Finally, a conductive target negates the electric even mode ($I_{e} = 0$, $I_{b} =-I_{a}$) and allows the source to be reduced to a single antenna (Fig. \ref{fig:ReducedHS}). Here, the image effect creates a virtual current loop by mirroring the electric current within the remaining element across the conductive plane \cite{Wong2016, Ang2020}.

\begin{figure}[t]
	\centering
	\includegraphics[width=0.45\textwidth]{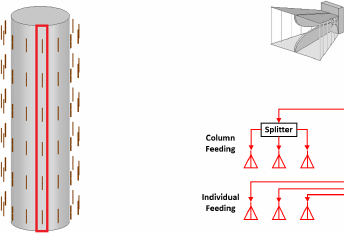}
	\caption{3D circular cloak model with element column boxed in red along with illustration of element feeding and control setups}
	\label{fig:3DLayout} 
\end{figure}

\section{3D Cloak Modeling \& Simulation}
An Ansys HFSS model of the 3D cloak problem is shown in Fig. \ref{fig:3DLayout}. As modeling and experiment design were conducted in parallel, model properties have been selected to best reflect their real-world equivalents. Since this 3D cloak represents a first-time proof-of-concept, design was performed conservatively and the resultant setup was obtained by extruding the 2D circular problem in Fig. \ref{fig:2DLayout} lengthwise.

Returning to Fig. \ref{fig:3DLayout}, a $753$ mm high conductive cylinder ($r=101.5$ mm radius) acts as the target and is illuminated at 1.2 GHz by an 1-18 GHz ridged horn antenna with its feed positioned $R_{t}=1424$ mm from target center. As the dimensions of the horn's experimental counterpart is proprietary, an equivalent ridged horn was generated using 3DS Antenna Magus and then imported into HFSS. The cloak itself is implemented using an $N=15$ element array of the reduced equivalent sources from Fig. \ref{fig:ReducedHS}. These are modeled as $50$ mm long dipole antennas positioned $w/2=34$ mm from the target's surface. To fully cover the target, 6 of these arrays are installed along the cylinder's length with a vertical spacing of $h_{s}=125.6$ mm between element centers. Subsequently for multiple arrays, $h = h_{s}$ in (\ref{eqn:Mdm}) and (\ref{eqn:Edm}).

\begin{figure*}[t]
	\centering
	\includegraphics[width=6.5in]{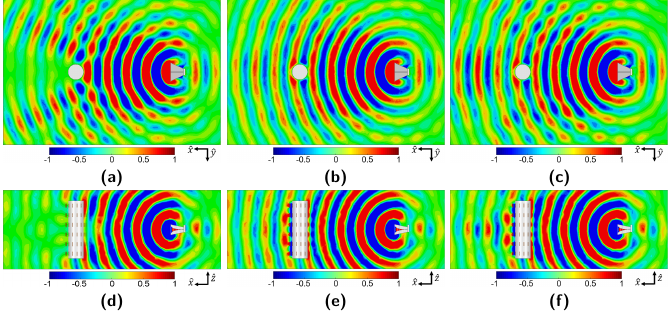}
	\begin{subfigure}{0em}
		\phantomsubcaption{}
		\label{fig:HFSSEFieldOFF}
	\end{subfigure}
	\begin{subfigure}{0em}
		\phantomsubcaption{}
		\label{fig:HFSSEFieldONInd}
	\end{subfigure}
	\begin{subfigure}{0em}
		\phantomsubcaption{}
		\label{fig:HFSSEFieldONCol}
	\end{subfigure}
	\begin{subfigure}{0em}
		\phantomsubcaption{}
		\label{fig:HFSSHFieldOFF}
	\end{subfigure}
	\begin{subfigure}{0em}
		\phantomsubcaption{}
		\label{fig:HFSSHFieldONInd}
	\end{subfigure}
	\begin{subfigure}{0em}
		\phantomsubcaption{}
		\label{fig:HFSSHFieldONCol}
	\end{subfigure}
	\caption{Simulated normalized, real total fields: (a) E-fields, cloak OFF (b) E-fields, cloak ON - individual feeding (c) E-fields, cloak ON - column feeding (d) H-fields, cloak OFF (e) H-fields, cloak ON - individual feeding (f) H-fields, cloak ON - column feeding}
	\label{fig:HFSSField} 
\end{figure*}

\subsection{Simulation setup \& element weighting}
Full-wave electromagnetic simulation (via Ansys HFSS) were then used to model fields and initially validate cloak performance. This analysis was set up by feeding both the source horn and cloak elements within the 3D model with current sources while the simulation volume was surrounded by PML (Perfectly Matched Layer) absorptive boundary conditions to enforce free-space conditions.

Due to their 2D nature, previous conceptual and experimental cloak designs were illuminated by simple planar and cylindrical incident wavefronts. This made array configuration straightforward as the incident fields, and cloak weights, could be calculated by simply substituting the location of each prospective element into the appropriate incident wave equation. Unfortunately, this method is not as ideal for more complicated scenarios such as illumination by a complex wavefront (i.e. ridged horn) or by multiple incident waves. One solution is to calculate weights using simulated incident fields. To accomplish this, the simulation is run without the target and cloak array present, leaving only the incident wavefront. These fields can then be directly sampled at the would be element locations \cite{Ang2020}.

\subsubsection{Element Feeding}
In an ideal setup, the input of each cloak element would be individually controlled. While this is easy to achieve within a simulated model, a real-world setup requires the means to control 90 separate signal channels which can be challenging to implement. One method to simplify feeding and control is to weight each vertical column of elements with the same input (Fig. \ref{fig:3DLayout}); preferably derived from field samples at or near mid-target height. For this initial proof-of-concept, such a setup is reasonable since the target is within the source horn's far-field and fields are sampled along a centered, radial cross-section. Within this region, the incident field is expected to be planar enough such that elevation-wise spherical wave phase differences will not significantly impact cloaking/scattered field matching. To verify this, in lieu of experimentation, both an ``individual fed" and a ``column fed" model (Fig. \ref{fig:3DLayout}) will be simulated and compared. For the latter, the elements are fed using weights based on incident field samples from the cylinder's mid-height.

\subsection{Simulated Field-Patterns}
Fig. \ref{fig:HFSSField} shows the normalized real Cloak OFF and Cloak ON electric (horizontal) and magnetic (vertical) field plots. Such field plots provide an easy means to verify cloak performance qualitatively while using real components accounts for both magnitude and phase variations. In the Cloak OFF case (Fig. \ref{fig:HFSSEFieldOFF} and \ref{fig:HFSSHFieldOFF}), with the array powered off, scattering to the front and sides of the cylinder can be observed along with the formation of a shadow region at the rear of the target. Fig. \ref{fig:HFSSEFieldONInd} and \ref{fig:HFSSHFieldONInd} then show the total fields when the cloak array inputs are configured individually. Here, it can be seen that scattering and shadowing have been minimized and the incident wave restored both horizontally and vertically. Column feeding the elements results in a similar horizontal E-field (Fig. \ref{fig:HFSSEFieldONCol}) distribution as the previous individually-fed setup. However from Fig. \ref{fig:HFSSHFieldONCol}, elevation mismatches and potential edge scattering reduces the effectiveness of scattering cancellation near the top and bottom of the cylinder. 

\begin{figure}[b]
	\centering
	\includegraphics[width=0.48\textwidth]{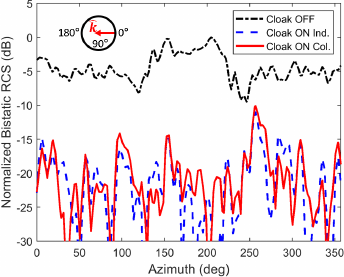}
	\caption{Simulated bistatic radar cross-section patterns}
	\label{fig:HFSSBRCS} 
\end{figure}

\begin{figure*}[t]
	\centering
	\includegraphics[width=6.5in]{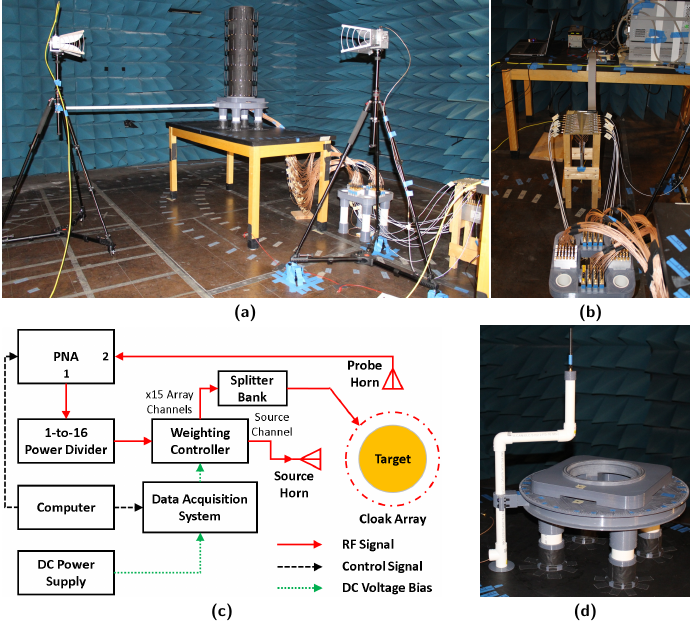}
	\begin{subfigure}{0em}
		\phantomsubcaption{}
		\label{fig:3DExperimentBistatic}
	\end{subfigure}
	\begin{subfigure}{0em}
		\phantomsubcaption{}
		\label{fig:3DExperimentController}
	\end{subfigure}
	\begin{subfigure}{0em}
		\phantomsubcaption{}
		\label{fig:3DExperimentFlow}
	\end{subfigure}
	\begin{subfigure}{0em}
		\phantomsubcaption{}
		\label{fig:3DExperimentNFProbe}
	\end{subfigure}
	\caption{3D cloak experiment: (a) bistatic apparatus (b) source/cloak control setup (c) system design (d) incident field sampling probe}
	\label{fig:3DExperiment} 
\end{figure*}

\subsection{Simulated Scattering Cross-Section}
To provide an initial quantitative measure of performance, the normalized bistatic radar cross-section (RCS) patterns for the Cloak OFF as well as the individual (Ind.) and column-fed (Col.) Cloak ON cases are plotted in Fig. \ref{fig:HFSSBRCS}. Due to the use of an illuminating horn and the physical size of the simulation (and mesh), it is not possible to use HFSS' built-in RCS calculators while field samples taken from far-field ranges contain excessive noise. Instead, bistatic RCS patterns are calculated from near-field samples taken along a $b=270.9$ mm, E-plane radial track. These are then extrapolated to the far zone using a cylindrical harmonic expansion method, designed for near-field ranges, detailed in \cite{Brown1961, Selvanayagam2013, Ang2020}. 

Overall, both patterns show that cloaking is achieved over the entire azimuth with an average suppression of 18.4 dB and 17 dB for the individually and column fed setups respectively. In general, these differences in average suppression can be attributed to the individually fed Cloak ON pattern having more nulls and higher maximum suppression (55.5 dB versus 41.8 dB for column fed). Furthermore, with the exception of the peaks at $255^{\circ}$, both Cloak ON patterns appear to stabilize to a consistent baseline ($\approx-20$-25 dB) regardless of the corresponding Cloak OFF value. This behavior has been seen in previous conceptual designs \cite{Ang2018} where areas of high Cloak OFF scattering (i.e. shadow region, $\approx180^{\circ}$) see more suppression than low scattering areas. Based on these RCS patterns, and the previous E-plane field plots, column feeding provides sufficient performance for centered, azimuthal scans such that this setup can be used for experimentation. 

\section{Experimental Validation}
The experimental apparatus used to validate real-world 3D cloaking performance can be seen pictured in Fig. \ref{fig:3DExperiment}. Here, an iron stovepipe serves as the target ($h=753$ mm, $r=101.5$ mm) and sits upon a PVC/3D printed (eSUN PLA+, Modix Big-60 3D printer) support structure (Fig. \ref{fig:3DExperimentBistatic}). A pair of A-Info L-10180-SF 1-18 GHz ridged horn antennas are then used to illuminate the target and measure the resultant bi-static radar cross-section (RCS). These are mounted on tripods, sighted to mid-target height, with the source horn fixed with its feed approximately $R_{t}=1424$ mm from target center. Conversely, the tripod supporting the measurement (probe) horn is wheeled and connected to a rotating segment of the support structure via a section of 1/2 in PVC pipe. This allows the probe horn to scan the target azimuthally while remaining at a (feed referenced) fixed radius of $R_{r}=1465$ mm. Additionally to avoid collisions between the measurement tripod and the rest of the apparatus, the resultant scan setup has an azimuthal range of $30-315^{\circ}$. Lastly, the cloak array is made up of 2.45 GHz whip dipole antennas (Microchip Technology RN-SMA-4) mounted via SMA bulkheads into holes drilled into the target. Here, the positioning and spacing of the elements is designed to match that of the HFSS model ($N=15$, 6 rows, $h_{0}=125.6$ mm, $w/2 = 34$ mm). 

The cloak elements and the source horn are fed and weighted by the control setup in Fig. \ref{fig:3DExperimentController}; the workings of which can be better understood by examining the general system block diagram from Fig. \ref{fig:3DExperimentFlow}. Here, an Agilent E8361C Network Analyzer (PNA) is responsible for both conducting pattern measurements and acting as a power source for the source and cloak antennas. This latter functionality is facilitated by connecting the PNA's Port 1 to a 16-channel controller which allows the inputs of the source antenna and of the 15 cloak element columns to be magnitude and phase scaled with respect to each other (Fig. \ref{fig:3DExperimentController}). Specifically, the controller operates by first using a 16-port Mini-Circuits ZC16PD-24-S+ splitter to divide the initial, Port 1 signal. To conduct magnitude and phase scaling (weighting), each resultant channel is equipped with a pair of voltage controlled attenuators (Mini-Circuits EVA-1500) and phase shifters (Mini-Circuits SPHSA-152). Biasing voltages for these components subsequently originate from a Kikusui PMC18-2A DC power supply and distributed using a pair of MATLAB-controlled data acquisition systems (Measurement Computing USB-3105). One of these 16 channels (source) is then directly routed to the source horn while the remaining 15 cloak channels are further divided by a 15-unit bank of 6-port splinters (Mini-Circuits ZB6PD-17-S), allowing them to feed their respective element columns (See Fig. \ref{fig:3DExperimentController}). Port 2 of the PNA is then connected to the probe horn to enable RCS pattern measurements. Lastly to reduce spurious reflections, the experimental setup is situated within an anechoic chamber with absorptive foam cones mounted along the chamber's ceiling and walls. 

\subsection{Experimental Weighting}

\begin{figure}[t]
\centering
\includegraphics[width=0.48\textwidth]{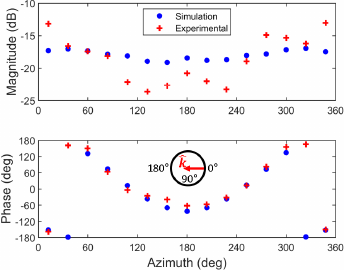}
\caption{Simulation and experiment-based magnetic current ($\mathbf{M}_{c}$) cloaking weights}
\label{fig:ExpWeights} 
\end{figure}

During experimentation, two approaches to element configuration were explored. The first ``simulation-based" approach fed the elements using the previous HFSS column fed weights. These had been phase and magnitude scaled to account for feeding differences between the source horn and cloak elements. However, this direct use of simulated incident fields also implies that it would also be possible to configure the cloak using real-world measurements. To test this hypothesis, a second ``experiment-based" set of weights were deduced using the probe in Fig. \ref{fig:3DExperimentNFProbe}. This would be attached to the scan base, replacing the regular probe boom, and used to sample the incident field at each element's azimuth angle along cylinder mid-height. Such a technique has the advantage of enabling cloak configuration without a-priori field knowledge but requires the removal of the target and cloak from the base.

Both sets of cloaking weights, in the form of magnetic currents ($\mathbf{M}_{c}$), can then be seen plotted in Fig. \ref{fig:ExpWeights} and show little phase variation; implying that the wavefront geometry from the approximated simulated horn is comparable to its physical counterpart. However, a variance of 5 dB can be seen in the experimental weight magnitudes, particularly at the back ($~180^{\circ}$) of the cylinder. This is likely due to the probe picking up any apparatus specific variances such as from scattering off the chamber floor, equipment, and support structures. In contrast, this excess clutter would not be present in the simulated HFSS environment.

\subsection{Experimental Cloak Only Fields}
After weighting the cloak elements, the probe antenna was azimuthally swept, at $5^{\circ}$ increments, to obtain ``Incident" , ``Cloak OFF" , and ``Cloak ON" patterns in the form of S-parameters. Specifically, the Incident pattern ($\mathbf{S}_{21,i}$) was measured with the cylinder (and cloak) removed from its support structure while the total Cloak OFF ($\mathbf{S}_{21,t_{OFF}}$) and ON ($\mathbf{S}_{21,t_{ON}}$) fields were sampled with the cylinder in place. It would then be possible to isolate the Cloak OFF ($\mathbf{S}_{21,s_{OFF}}$) and ON ($\mathbf{S}_{21,s_{ON}}$) scattering along with the fields radiated by the cloak array ($\mathbf{S}_{21,c}$):

\begin{gather}
\mathbf{S}_{21,s_{OFF}}=\mathbf{S}_{21,t_{OFF}}-\mathbf{S}_{21,i}\\
\mathbf{S}_{21,s_{ON}}=\mathbf{S}_{21,t_{ON}}-\mathbf{S}_{21,i}\\
\mathbf{S}_{21,c}=\mathbf{S}_{21,t_{ON}}-\mathbf{S}_{21,t_{OFF}}
\end{gather}

\begin{figure}[t]
\centering
\includegraphics[width=0.48\textwidth]{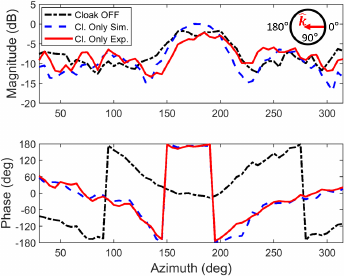}
\caption{Comparison of scattered and cloak-radiated fields}
\label{fig:ExpCloakOnly} 
\end{figure}

Fig. \ref{fig:ExpCloakOnly} compares the phase and (normalized) magnitude of these ``Cloak Only" fields against the Cloak OFF scattering. Overall, magnitude variances between the cloak and scattered fields were found to be 1.25 and 0.31 dB for the simulation and experimental-based configurations respectively. That is, better matching is observed in the experimental configuration, particularly at the shadow region ($160^{\circ}$-$195^{\circ}$). To facilitate scattering cancellation, both sets of cloak fields maintain a constant phase offset near $180^{\circ}$ with respect to their Cloak OFF scattered counterpart. Taking this phase inversion into account, the simulation and experimental configurations possess average phase variances of $16^{\circ}$ and $12^{\circ}$ respectively.

\subsection{Experimental Scattering Cross-Section}

As the experimental measurements are sampled in the far-field, the radar range equation can be used to deduce the bistatic RCS ($\sigma$) patterns \cite{Balanis2012b, Soric2013}:

\begin{equation}
\left|\mathbf{S}_{21,s}\right|^{2} = \frac{P_{r}}{P_{t}} = \frac{G_{t}G{r}\lambda^{2}_{0}\sigma}{(4\pi)^{3}R_{t}^{2}R_{r}^{2}}
\end{equation}

where $P$, $G$, and $\lambda_{0}$ refer to the receiving ($r$)/transmitting ($t$) power, horn gain, and operating wavelength respectively. 

\begin{figure}[t]
\centering
\includegraphics[width=0.48\textwidth]{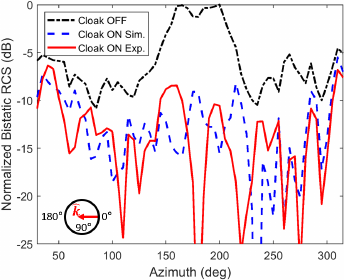}
\caption{Experimental bistatic radar cross-section patterns}
\label{fig:ExpBRCS} 
\end{figure}

These RCS patterns are then plotted in Fig. \ref{fig:ExpBRCS} which show that both configurations achieve scattering suppression across the entire measured azimuth. Specifically, simulation and experiment-based patterns show scattering suppression averages of 7.7 and 8.3 dB respectively with global maxima of 37.5 and 34.2 dB. Here, it can be seen that the experimental-based Cloak ON pattern, contains multiple steep local minima within the shadow region. It is likely that these are responsible for the respective elevation in average suppression.

\begin{figure}[t]
\centering
\includegraphics[width=0.48\textwidth]{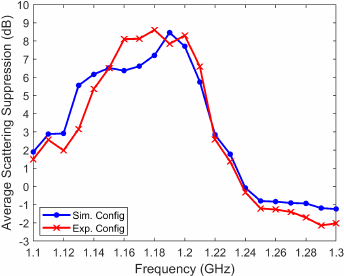}
\caption{Experimental single weight sweep at 1.2 GHz}
\label{fig:ExpBW} 
\end{figure}

\begin{figure}[t]
\centering
\includegraphics[width=0.48\textwidth]{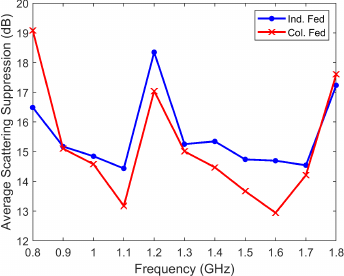}
\caption{Frequency hopping simulations}
\label{fig:SimFS} 
\end{figure}

\section{Multi-Frequency Capabilities}
Present research has primarily focused on demonstrating the feasibility of active cloaking on various targets along with refining the cloak's design and implementation. As such, prototypes to date had been designed specifically for single frequency operation. Although narrowband devices can still be useful, their practicality is greatly limited. Since this research aims to advance the overall versatility of active cloaking, it is appropriate to include a conceptual investigation into present and potential multi-frequency capabilities. Theoretically, an active cloak should not be subject to any frequency-based limitations. That is, cloaking should be achievable as long as the appropriate $\mathbf{J}_{c}$ and $\mathbf{M}_{c}$ can be generated. Therefore, it is mainly the cloak's design and operating environment which determines its behavior over other frequencies. 

The discussion on multi-frequency can be divided into two sections: single weight bandwidth and frequency hopping. The former concerns the behavior observed at other frequencies while the cloak is configured to operate at a specific frequency (1.2 GHz). On the other hand, frequency hopping or reconfigurability refers to the ability of a cloak to shift its operating frequency, generally without physical modification. Both of these characteristics can be investigated using existing simulated or experimental results and provide potential directions for development which can be built upon in future designs and experiments. 

\subsection{Single Weight Bandwidth: Experimental}
Fig. \ref{fig:ExpBW} displays the experimental average suppression observed between 1.1 GHz to 1.3 GHz when the cloak is configured for 1.2 GHz. Here, both the simulated and experimental weighted cloaks posses similar functional bandwidths of 13.9\% and 13.2\% respectively as well as 3 dB bandwidths of 7.4\% and 6.2\%. Since the experimental configuration is based on actual incident field measurements, it can be seen that the experimental-based configuration maintains near peak suppression over a wider frequency range but experiences a rapid drop-off past 1.15 GHz. One possible explanation is that the increased precision of the weights may lower tolerance to cloak/scattered field mismatches. Alternatively, this behavior may be caused by the weights accounting for apparatus specific variances. These may frequency scale differently causing or amplifying field mismatches. Lastly, the experimental configuration achieves peak performance at 1.18 GHz (8.6 dB) while the simulated weights result in a peak frequency of 1.19 GHz (8.45 dB).

\subsection{Frequency Hopping: Simulation}
One advantage of operating non-resonant cloak elements is that it is easy to re-weight the cloak to operate at another center frequency. This is demonstrated in Fig. \ref{fig:SimFS} in which the simulated model from Fig. \ref{fig:3DLayout} is reconfigured to operate at center frequencies spanning from 0.8 to 1.8 GHz. Here, both feeding configurations are shown to be capable of maintaining similar suppression baselines above 13 dB. Based on previous experience \cite{Ang2020}, there are two main issues which may be encountered when reconfiguring the cloak's operating frequency. First, is the occurrence of mode scattering when operating near the cloak array's resonant frequency (2.45 GHz). In this case, a second resonant-based scattering mode arises in addition to the existing structural scattered field \cite{Lynch2004, Knott2004}. This additional field component is much more challenging than conventional (structural) scattering to determine and may require computational solutions. However if not accounted for during element weighting, the cloak field will no longer match with the scattered field, degrading cloak performance. The easiest method of mitigating this issue is though design-based solutions. These include selecting elements which are not resonant under the expected operating frequency range or intentionally impedance mismatching a resonant antenna. Furthermore, in addition to reducing antenna mode scattering, these measures also minimize inter-element coupling. The second frequency hopping issue is that at much higher frequencies, array aliasing and element current phase inversion may reduce performance \cite{Ang2020}. The former occurs when the array spacing exceeds half the operational wavelength. This prevents the cloak from accurately reproducing the scattered field, leading to imperfect matching. In addition, current phase inversion occurs when the electrical size of the elements becomes too large with respect to the operating wavelength. Here, the currents within the elements may phase invert and cancel out. Consequently, these effects can generally be mitigated by keeping element length (for dipoles/monopoles) and spacing smaller than half a wavelength at the highest expected operating frequency.

The experimental cloak in this paper borrows its design and components from the quasi-2D experiment in \cite{Ang2020}. As such, operational bandwidth is limited primarily by the cloak controller which is only capable of reliably magnitude and phase scaling at 1.2 GHz. While it is necessary to rebuild the controller to experimentally validate frequency hopping, this also opens up potential future design directions. For instance, a more capable controller may be able to maintain weighting across multiple frequencies. Such a ``composite" weighting approach may enable a cloak to suppress scattering at multiple individual frequencies and/or across a frequency range.

\section{Conclusion}
This paper presents the first steps in the development and demonstration of active interior cloaking on a free-space, 3D target. As an initial proof of concept, a conservative design approach was adopted in which the 3D cloaking layout was derived by extruding a previous 2D cloak problem height-wise. In this setup, a conductive circular cylinder served as the target and illumination was provided at $1.2$ GHz by a ridged horn antenna. The designed cloak then took the form of 6, 15 element dipole arrays covering the cylinder lengthwise. Following initial validation by full-wave simulations, cloak performance was assessed under real-world conditions using an experimental bistatic measurement setup. Due to constraints imposed by available control equipment, the experimental elements needed to be fed in columns. Despite operating in this simplified format, the prototype cloak achieved scattering suppression over all measured points when scanned azimuthally across the target's mid-height; 7.7 - 8.3 dB average suppression. Although active cloaking is not theoretically subject to frequency limitations, the array and controller properties limited the cloak's functional and 3 dB bandwidths to 13-14 \% and 6-7.5 \% respectively when configured for a center frequency of 1.2 GHz. Lastly, although the present experimental cloak controller was only capable of operating at 1.2 GHz, full-wave simulations were successfully used to demonstrate that the cloak could be re-weighted for single frequency operation across 0.8 - 1.8 GHz without physical alteration.

Immediate future tasks are expected to focus on expanding the measurement capabilities of the apparatus and demonstrating cloaking on more complex 3D targets. In the case of the former, the size and bulk of the existing experiment has limited both illumination and RCS measurement to a single azimuthal plane. As such, a redesigned probe antenna should ideally be capable of altering its elevation such that scattering suppression can be verified over a 3D sphere. Likewise, an improved source antenna would be capable of illuminating the target at various elevation angles to better reflect real-world conditions. However, factoring elevation into illumination and measurement may require the cloak itself to be redesigned to enable individual element feeding, for height-wise field variations, and to accommodate potential edge scattering as well as multiple incident polarizations; elements with two orthogonal linear responses can work with all polarizations.

Longer term goals are then aimed at expanding and exploring additional cloak capabilities in general. The first of these is to begin integration of an incident field estimator, such as that in \cite{Ang2021}. Here, cloak elements can be re-purposed to operate as sensors to measure the total field, in close proximity to the target, which could then be used by the algorithm to deduce the incident field. Subsequently, pairing an estimator with a controller would then allow an active cloak to autonomously sense and adapt to an unknown and dynamic electromagnetic environment without outside intervention nor a-priori field knowledge. Another important capability for a practical cloak to have is the ability to accommodate wide-band or frequency-variant signals. To explore and develop this capacity, the control board must first be redesigned to allow weighting over a wide frequency range. This would enable experimental demonstration of frequency hopping where a cloak could be reconfigured to operate at another single frequency. More advanced controller designs may then have the ability to enforce element weighting across a frequency range. This would open up the possibility of multi-frequency devices capable of cloaking across multiple individual frequencies or a frequency range. 

\printbibliography[title={References}]

@PREAMBLE{
 "\providecommand{\noopsort}[1]{}" 
 # "\providecommand{\singleletter}[1]{#1}%" 
}

@article {Schurig2006,
	author = {Schurig, D. and Mock, J. J. and Justice, B. J. and Cummer, S. A. and Pendry, J. B. and Starr, A. F. and Smith, D. R.},
	title = {Metamaterial Electromagnetic Cloak at Microwave Frequencies},
	volume = {314},
	number = {5801},
	pages = {977--980},
	year = {2006},
	publisher = {American Association for the Advancement of Science},
	journal = {Science}
}

@article{Kim2016,
	author = {Yongjune Kim and Ilsung Seo and Il-Suek Koh and Yongshik Lee},
	journal = {Opt. Express},
	number = {20},
	pages = {22708--22717},
	publisher = {OSA},
	title = {Design method for broadband free-space electromagnetic cloak based on isotropic material for size reduction and enhanced invisibility},
	volume = {24},
	year = {2016}
}

@article{Ma2013,
	title = {First experimental demonstration of an isotropic electromagnetic cloak with strict conformal mapping},
	author = {Yungui Ma and Yichao Liu and Lu Lan and Tiantian Wu and Wei Jiang and C. K. Ong and Sailing He},
	journal = {Scientific Reports},
	volume = {3},
	number = {1},
	year = {2013}
}

@article{Chen2012,
	author = {Chen, Hongsheng and Zheng, Bin},
	title = {Broadband polygonal invisibility cloak for visible light},
	volume = {2},
	number = {255},
	year = {2012},
	journal = {Scientific Reports}
}

@article{Rainwater2012,
	author={D. Rainwater and A. Kerkhoff and K. Melin and J. C. Soric and G. Moreno and A. Al\`u},
	title={Experimental verification of three-dimensional plasmonic cloaking in free-space},
	journal={New Journal of Physics},
	volume={14},
	number={1},
	pages={013054},
	year={2012}
}

@article{Vitiello2016,
	title = {Waveguide Characterization of S-Band Microwave Mantle Cloaks for Dielectric and Conducting Objects},
	author = {Vitiello, Antonino and Moccia, Massimo and Papari, Gian Paolo and D’Alterio, Giuliana and Vitiello, Roberto and Galdi, Vincenzo and Andreone, Antonello},
	journal = {Scientific Reports},
	volume = {6},
	year = {2016},
	publisher = {Scientific Reports}
}

@article{Qin2018,
	title = {Mantle Cloaks Based on the Frequency Selective Metasurfaces Designed by Bayesian Optimization},
	author = {F. F. Qin and Z. Z. Liu and Q. Zhang and H. Zhang and J. J. Xiao},
	journal = {Scientific Reports},
	volume = {8},
	number = {14033},
	year = {2018}
}

@article{Fleury2015,
	title = {Invisibility and Cloaking: {o}rigins, Present, and Future Perspectives},
	author = {Fleury, Romain and Monticone, Francesco and Al\`u, Andrea},
	journal = {Phys. Rev. Applied},
	volume = {4},
	issue = {3},
	pages = {037001},
	numpages = {20},
	year = {2015},
	publisher = {American Physical Society},
}

@article{Monticone2013,
	title = {Do Cloaked Objects Really Scatter Less?},
	author = {Monticone, Francesco and Al\`u, Andrea},
	journal = {Phys. Rev. X},
	volume = {3},
	number = {4},
	pages = {041005},
	year = {2013},
	publisher = {American Physical Society}
}

@article{Zeng2014,
	author = {Zeng, Chao and Liu, Xueming and Wang, Guoxi},
	journal = {Scientific Reports},
	number = {5763},
	title = {Electrically tunable graphene plasmonic quasicrystal metasurfaces for transformation optics},
	volume = {4},
	year = {2014}
}

@article{Hosseininejad2019,
	author = {Hosseininejad, Seyed Ehsan and Rouhi, Kasra and Neshat, Mohammad and Faraji-Dana, Reza and Cabellos-Aparicio, Albert and Abadal, Sergi and Alarc\'on, Eduard},
	journal = {Scientific Reports},
	number = {2868},
	title = {Reprogrammable Graphene-based Metasurface Mirror with Adaptive Focal Point for THz Imaging},
	volume = {9},
	issue = {1},
	year = {2019}
}

@article{Meng2017,
	author = {Meng, Qinglong and Zhong, Zheqiang and Zhang, Bin},
	journal = {Scientific Reports},
	number = {45708},
	title = {Hybrid three-dimensional dual- and broadband optically tunable terahertz metamaterials},
	volume = {7},
	year = {2017}
}

@article{Yang2016b,
	author = {Yang, Huanhuan and Cao, Xiangyu and Yang, Fan and Gao, Jun and Xu, Shenheng and Li, Maokun and Chen, Xibi and Zhao, Yi and Zheng, Yuejun and Li, Sijia},
	journal = {Scientific Reports},
	number = {35692},
	title = {A programmable metasurface with dynamic polarization, scattering and focusing control},
	volume = {6},
	year = {2016}
}

@article{Chen2013,
	title = {Broadening the Cloaking Bandwidth with Non-{F}oster Metasurfaces},
	author = {Chen, Pai-Yen and Argyropoulos, Christos and Al\`u, Andrea},
	journal = {Phys. Rev. Lett.},
	volume = {111},
	number = {23},
	pages = {233001},
	year = {2013},
	publisher = {American Physical Society}
}

@article{Liu2014,
	author = {Shuo Liu and He-Xiu Xu and Hao Chi Zhang and Tie Jun Cui},
	journal = {Opt. Express},
	number = {11},
	pages = {13403--13417},
	publisher = {OSA},
	title = {Tunable ultrathin mantle cloak via varactor-diode-loaded metasurface},
	volume = {22},
	year = {2014}
}

@article{Farhat2013,
	author = {Mohamed Farhat and Carsten Rockstuhl and Hakan Ba\u{g}c{\i}},
	journal = {Opt. Express},
	number = {10},
	pages = {12592--12603},
	publisher = {OSA},
	title = {A {3D} tunable and multi-frequency graphene plasmonic cloak},
	volume = {21},
	year = {2013}
}

@article{Zhu2013,
	author = {Xuefeng Zhu and Liang Feng and Peng Zhang and Xiaobo Yin and Xiang Zhang},
	journal = {Opt. Lett.},
	number = {15},
	pages = {2821--2824},
	publisher = {OSA},
	title = {One-way invisible cloak using parity-time symmetric transformation optics},
	volume = {38},
	year = {2013}
}

@ARTICLE{Zhu2014,
	author={H. L. Zhu and X. H. Liu and S. W. Cheung and T. I. Yuk},
	journal={IEEE Transactions on Antennas and Propagation},
	title="{Frequency-Reconfigurable Antenna Using Metasurface}",
	year={2014},
	volume={62},
	number={1},
	pages={80-85}
}

@article{Miller2006,
	author = {David A. B. Miller},
	journal = {Opt. Express},
	number = {25},
	pages = {12457--12466},
	publisher = {OSA},
	title = {On perfect cloaking},
	volume = {14},
	year = {2006}
}

@article{Vasquez2009,
	title = {Active Exterior Cloaking for the {2D} {L}aplace and {H}elmholtz Equations},
	author = {Vasquez, Fernando Guevara and Milton, Graeme W. and Onofrei, Daniel},
	journal = {Phys. Rev. Lett.},
	volume = {103},
	number = {7},
	pages = {073901},
	year = {2009},
	publisher = {American Physical Society}
}

@ARTICLE{Selvanayagam2012,
	author={M. Selvanayagam and G. V. Eleftheriades},
	journal={IEEE Antennas and Wireless Propagation Letters},
	title={An Active Electromagnetic Cloak Using the Equivalence Principle},
	year={2012},
	volume={11},
	number={},
	pages={1226-1229}
}

@article{Zheng2010b,
	title = {Exterior optical cloaking and illusions by using active sources: {a} boundary element perspective},
	author = {Zheng, H. H. and Xiao, J. J. and Lai, Y. and Chan, C. T.},
	journal = {Phys. Rev. B},
	volume = {81},
	number = {19},
	pages = {195116},
	year = {2010},
	publisher = {American Physical Society}
}

@article{Du2012,
	author = {Junjie Du and Shiyang Liu and Zhifang Lin},
	journal = {Opt. Express},
	number = {8},
	pages = {8608--8617},
	publisher = {OSA},
	title = {Broadband optical cloak and illusion created by the low order active sources},
	volume = {20},
	year = {2012}
}

@article{Wong2018,
	title={Active Huygens' box: metasurface-enabled arbitrary electromagnetic wave generation inside a cavity},
	author={Wong, Alex MH and Eleftheriades, George V},
	journal={arXiv preprint arXiv:1810.05998},
	year={2018}
}

@article{Selvanayagam2013,
	title = {Experimental Demonstration of Active Electromagnetic Cloaking},
	author = {Selvanayagam, Michael and Eleftheriades, George V.},
	journal = {Phys. Rev. X},
	volume = {3},
	number = {4},
	pages = {041011},
	year = {2013},
	publisher = {American Physical Society}
}

@INPROCEEDINGS{Ang2018,
	author={P. Ang and G. V. Eleftheriades},
	booktitle={2018 IEEE/MTT-S International Microwave Symposium - IMS},
	title={Active {H}uygens' Cloaks for Arbitrary Metallic Polygonal Cylinders},
	year={2018},
	volume={},
	number={},
	organization={IEEE},
	pages={337-340}
}

@INPROCEEDINGS{Ang2018a,
	author={P. Ang and G. V. Eleftheriades},
	booktitle={2018 IEEE International Symposium on Antennas and Propagation \& USNC/URSI National Radio Science Meeting},
	title={Active Surface Cloaking with Patch Antennas},
	year={2018},
	volume={},
	number={},
	pages={911--912},
	organization={IEEE}
}

@INPROCEEDINGS{Ang2020a,
	author={P. Ang and G. V. Eleftheriades},
	booktitle={2020 IEEE International Symposium on Antennas and Propagation},
	title={Active Cloaking Without A Priori Incident Field Knowledge},
	year={2020},
	volume={},
	number={},
	organization={IEEE}
}

@article{Ang2020,
	author = {Paris Ang and George V. Eleftheriades},
	year = {2020},
	title = {Active Cloaking of a Non-Uniform Scatterer},
	volume = {10},
	number = {2021},
	pages={1--11},
	journal = {Scientific Reports},
	publisher={Nature Publishing Group}
}

@article{Ang2021,
	title={Incident-Field Estimation for Active Cloaking},
	author={Ang, Paris and Eleftheriades, George V},
	journal={Physical Review Applied},
	volume={16},
	number={6},
	pages={064005},
	year={2021},
	publisher={APS}
}

@inproceedings{Ang2021a,
	title={Active Cloaking with an Incident-Field Estimation Algorithm},
	author={Ang, Paris and Eleftheriades, George V},
	booktitle={2021 IEEE MTT-S International Microwave Symposium (IMS)},
	pages={478--481},
	year={2021},
	organization={IEEE}
}

@INPROCEEDINGS{Wong2016,
	author={A. M. H. Wong and G. V. Eleftheriades},
	booktitle={2016 18th Mediterranean Electrotechnical Conference (MELECON)},
	title={Active {H}uygens' metasurfaces for {RF} waveform synthesis in a cavity},
	year={2016},
	volume={},
	number={},
	pages={1-5},
	organization={IEEE}
}

@book{Harrington2001,
	title={{T}ime-{H}armonic {E}lectromagnetic {F}ields},
	author={Harrington, R.F.},
	isbn={978-0-471-20806-8},
	year={2001},
	publisher={Wiley-IEEE Press}
}

@INPROCEEDINGS{Ang2019,
	author={P. Ang and G. V. Eleftheriades},
	booktitle={2019 IEEE/MTT-S International Microwave Symposium - IMS},
	title={Experimental Active Cloaking of a Metallic Polygonal Cylinder},
	year={2019},
	volume={},
	number={},
	organization={IEEE},
	pages={}
}

@book{Balanis2012b,
	title={{A}ntenna {T}heory: {A}nalysis and {D}esign},
	author={Balanis, C.A.},
	year={2012},
	publisher={Wiley}
}

@ARTICLE{Brown1961,
	author={J. {Brown} and E. V. {Jull}},
	journal={Proceedings of the IEE - Part B: Electronic and Communication Engineering},
	title={The prediction of aerial radiation patterns from near-field measurements},
	year={1961},
	volume={108},
	number={42},
	pages={635-644}
}

@article{Soric2013,
	year = 2013,
	publisher = {{IOP} Publishing},
	volume = {15},
	number = {3},
	pages = {033037},
	author = {J. C. Soric and P. Y. Chen and A. Kerkhoff and D. Rainwater and K. Melin and A. Al{\`{u}}},
	title = {Demonstration of an ultralow profile cloak for scattering suppression of a finite-length rod in free space},
	journal = {New Journal of Physics}
}

@book{Lynch2004,
	title={Introduction to {RF} Stealth},
	author={Lynch, David},
	year={2014},
	publisher={Scitech Publishing Inc}
}

@book{Knott2004,
	title={Radar cross section},
	author={Knott, Eugene F and Schaeffer, John F and Tulley, Michael T},
	year={2004},
	publisher={SciTech Publishing}
}




\end{document}